\documentclass[prl,showclass, twocolumn,preprintnumbers,amsmath,amssymb]{revtex4}

\usepackage{graphicx}% Include figure files
\usepackage{dcolumn}% Align table columns on decimal point

\begin{document}

%\preprint{APS/123-QED}

\title{Fast random number generation\\ 
using 128 bit multimedia extension registers on Pentium class machines
}

\author{Borko D. Sto\v si\' c}
\email{borko@ufpe.br}
\affiliation{
Departamento de Estat\' \i stica e Inform\' atica, 
Universidade Federal Rural de Pernambuco,\\
Rua Dom Manoel de Medeiros s/n, Dois Irm\~ aos,
52171-900 Recife-PE, Brasil
}

\date{\today}% It is always \today, today,

\begin{abstract}
In this work it is shown how 128 bit SSE2 multimedia extension registers, present
in Pentium IV class 32 bit processors, may be used to generate random numbers at
several times greater speed then when regular general purpose registers are
used. In particular, a 128 bit algorithm is presented for the Marsaglia 
MWC1616 generator from the DIEHARD battery of random number generator tests,
and its performance is compared to that of the conventional approach.
\end{abstract}


\maketitle

Scientific computing has seen some impressive development over the last couple of 
decades, made possible by the exponential advent of cheap computer resources: 
what was doable a decade ago only on a supercomputer, today can be implemented on a laptop.
On the other hand, the appetites of scientists using computational techniques
have grown in parallel, so that algorithms used nowadays often require weeks 
(or months) of continuous processing, same as (orders of magnitude more modest) 
algorithms used a decade or two ago. 
While we can certainly expect the performance of processors
to continue improving (driven by general market requirements), 
it is a fact that during 
a single clock tick of a (currently common) 3GHz processor, the light traverses 
only 10cm - roughly twice the linear dimension of the actual processor chip
(this is actually an overestimate, as electrical signals propagate 
through matter at speeds lower then the speed of light in vacuum of 
$3\times 10^8$  m/s). Significant
further increase in clock times is therefore closely 
related to reducing the physical dimensions of the chip, 
and currently, parallel computing seems to be the most promising way out
of these physical limitations. 

In fact, the Intel processor manufacturer has already adopted this strategy \cite{intel}
some years ago, and most personal computers today already contain parallel
processing hardware capability. More precisely, 
starting with Pentium II processors, SIMD (Single Instruction Multiple Data) 
parallel computation on eight new internal 64 bit registers (called MM0-MM7) 
was introduced (this standard was named MMX, an abbreviation for Multimedia Extension). 
The concept was further improved in Pentium III processors by introducing another eight 
128 bit registers (called XMM0-XMM7) with a corresponding instruction set
(the standard was named SSE - Streaming SIMD Extensions), 
and in Pentium IV the SSE2 standard 
was implemented with an improved instruction set.
Finally, in the Pentium IV processor 3.40 GHz, supporting Hyper-
Threading Technology, SSE3 standard was introduced with additional
thirteen instructions.
The current SSE2/3 standard
works with packed data (two double precision, four single precision, 
or 16/8/4/2 integers of 8/16/32/64 bits each), where a single instruction is 
{\it simultaneously} executed on the packed
variables. When dealing with single precision floating point numbers or 32 bit
integers, this architecture yields (roughly) four times the performance of regular SISD
(Single Instruction Single Data) processing {\it on the same processor}.

It seems that the above developments have not been widely recognized by the
scientific community, while evidently they may prove crucial in a number of
situations where performance is critical. One such example is the (pseudo) random
number generation, which often proves to be the bottleneck in high performance
simulations (such as Monte Carlo, Simulated Annealing, Bootstrap, etc.) requiring
high precision, when high periodicity and low correlation of the random
number sequence is required.

In this work, a 128 bit algorithm is presented for the Marsaglia's MWC1616 \cite{diehard}
generator
with a period of roughly $2^{59}$, which passes all of the tests in the 
stringent DIEHARD suite of random number generator tests \cite{diehard,marsaglia1}.
Performance of this algorithm is then compared
with the conventional approach, both using 
function calls and C language macros.
First a brief overview of the MWC1616 generator is given,
followed by a naive but straightforward implementation in C, 
then a better macro version,
and finally the 128 bit parallel algorithm. 
Next, 
the performance testing results of the three algorithms are presented,
obtained on a Celeron 1.6GHz processor. Finally, the conclusions
are drawn.

The MWC1616 uniform (pseudo)random number generator \cite{diehard}
concatenates results 
of two 16 bit MWC (Multiple With Carry) generators \cite{marsaglia2}
to produce a 32 bit result.
The two 16 bit generators have the form
\begin{eqnarray}
x_{n+1}=&&\left[a\,x_n+c_n\right]\, mod\,\, 2^{16},\nonumber\\
y_{n+1}=&&\left[b\,y_n+d_n\right]\, mod\,\, 2^{16},
\label{one}
\end{eqnarray}
where $a$ and $b$ are multipliers (a good choice is $a=18000$ and $b=30903$,
a table of suggested values can be found in \cite{diehard}), $c_n$ and $d_n$ are
corresponding 16 bit overflow values (carry) resulting from 16 bit
multiplication at level $n$, 
and the symbol ``$mod$" indicates modulus operation.
The period of this generator is given by 
$\left(a\times2^{15}-1\right)\left(b\times2^{15}-1\right)$ \cite{diehard}, yielding over
$2^{59}\sim 6\times 10^{17}$ for the choice $a=18000$ and $b=30903$.
The generator is easily implemented in C by using unsigned long 32 bit integers
to store $x_n$ and $y_n$ in the low words and carry values $c_n$ and $d_n$ 
in the high words, using only two statements
\begin{verbatim}
x=a*(x&0xFFFF)+(x>>16);
y=b*(y&0xFFFF)+(y>>16);
\end{verbatim}
while the 32 bit integer to be returned
(which represents the member of the random sequence), 
is calculated by concatenating the two low words with the statement
\begin{verbatim}
(x<<16)+(y&0xFFFF);
\end{verbatim}

Finally, before running the generator  
one needs to initialize the two 16 bit seeds and the initial carry values, 
by choosing a value for 
$0<x_0\leq 2147483647$  and 
$0<y_0\leq 2147483647$ 
(2147483647 is decimal for 0x7FFFFFFF, 
the 16 bit seed is stored in the low word, 
and initial carry value in the high word).

The most straightforward full implementation in C programming language 
of the MWC1616 generator given by equation (\ref{one}), is given below
\begin{verbatim}
static unsigned long x=1, y=2;

void seed(unsigned long x0, unsigned long y0)
{
x=x0;
y=y0;
}

unsigned long MWC1616()
{
x=18000*(x&0xFFFF)+(x>>16);
y=30903*(y&0xFFFF)+(y>>16);
return (x<<16)+(y&0xFFFF);
}
\end{verbatim}
where the initial
seed values have been set to $x_0=1$, $y_0=2$ and $c_0=d_0=0$.
The seed() function should be called on initialization, 
before the first call to MWC1616(), to specify the origin of
the random number sequence.

While extremely simple and efficient 
(in comparison with many other random number generators), 
this implementation suffers from
the fact that each function call is accompanied by function prologue and
epilogue overhead (sequences of assembly instructions inserted by the
compiler on the beginning and the end, respectively, of any function call).
It is rather more efficient to replace the MWC1616() function call by a 
C language macro, as follows:
\begin{verbatim}
#define xnew	(x=(18000*(x&0xFFFF)+(x>>16)))
#define ynew	(y=(30903*(y&0xFFFF)+(y>>16)))
#define MWC1616	((xnew<<16)+(ynew&0xFFFF))
\end{verbatim}
This approach dispenses with the function call overhead
(the machine language opcode is literally inserted by the
compiler at all places where the macro is called),
while implementing exactly the same sequence of operations.
The function seed() need not be replaced with a macro, as it is
called only on rare occasions (normally only upon initialization).

In order to implement the MWC1616 generator in 128 bit arithmetic,
in what follows the SSE2 Pentium IV standard shall be used, 
as this is currently probably the most widespread situation,
and it seems that the SSE3 extensions found in newer processors 
do not provide any additional functionality relevant for the
current implementation.
Unfortunately, in order to implement the MWC1616 generator using the 
128 bit XMM registers and the SSE2 instruction set, one needs to resort 
to assembly language programming. The problem is that these are highly
specialized features of the processor, introduced mainly for the
purpose of multimedia streaming and high performance graphics
processing. Consequently, the high language compiler manufacturers
have not found an interest in incorporating these features in any of the
high level languages (such as C or Fortran), with the exception of the
Intel C compiler. As the current author had no access to this particular compiler
at the time of writing this paper, inline assembly code was used within the
Microsoft Visual Studio 6.0 environment, using the cl 32 bit C/C++ compiler. 
In order for this environment to recognize the SSE2
extensions together with the XMM register set, first the Intel SSE2 processor pack
had to be installed. It should be mentioned here that the use of the Intel
C compiler essentially requires identical effort in writing SSE2 code as assembly programming,
as the C language instructions have practically one-to-one correspondence
with their assembly language equivalents.

Although most personal computers nowadays fall into Pentium IV class,
before embarking on SSE2 assembly programming one should check for the
existence of these features on the processor to be used. This can be
accomplished (in assembly) by loading the eax register with value 1, 
issuing a CPUID instruction, and then examining the edx register: 
if the bit 26 is set, the processor supports SSE2 extensions.
The following function in C with inline assembly may be used to
implement this test (here the ``//" symbol indicates a comment from 
the current position to the end of the line):
\begin{verbatim}
int SSE2Available()
{
int available = 0;
_asm
  {
  mov eax, 1        //load eax register with 1
  cpuid             //issue CPUID
  shr edx, 26       //shift right 26 places
  and edx, 1        //mask out other bits
  mov [available], edx  //copy result
  };
return available;
}
\end{verbatim}
If the processor does not support SSE2 extensions the algorithm
described in the rest of this paper will not work.
It should be mentioned here that the check for support of 
earlier standards SSE and MMX may be performed by examining bits 25 and 23,
respectively of edx after CPUID instruction call, 
while support of the latest SSE3 standard is returned in bit 0
of the ecx register.

Let us now turn to the actual 128 bit 
implementation of the MWC1616 generator.
First, as the current implementation deals with fourfold parallel
instructions, one needs to initialize eight seeds (rather then only two).
Next, manipulation of registers is generally faster then operations
performed between registers and memory, so constant arrays are
first declared and initialized, to be loaded in scratch xmm registers
on startup. The following code snippet was used to declare and initialize
variable arrays and constants
\begin{verbatim}
static unsigned int	 
  x[4]={1,1,1,1},
  y[4]={2,2,2,2},
  r[4],
  mask[4]={0xFFFF,0xFFFF,0xFFFF,0xFFFF},
  mul1[4]={18000,18000,18000,18000},
  mul2[4]={30903,30903,30903,30903};

_asm
  {
  movdqa xmm0, x    //load array x into xmm0
  movdqa xmm1, y    //load array y into xmm1
  movdqa xmm5,mask  //load mask
  movdqa xmm6,mul1  //load first multiplier
  movdqa xmm7,mul2  //load second multiplier
}
\end{verbatim}
where the SSE2 instruction ``movdqa" loads values of variable arrays x and y
into registers xmm0, xmm1, and constant arrays mask, mul1 and mul2 
into registers  xmm5, xmm6 and xmm7, respectively, for posterior use. 
Note that the arrays x and y have been
initialized with same seeds for all four indices for testing purposes (to verify
whether all the four adjacent 32 bit blocks yield the same result), in a real application 
all four pairs should be initialized to distinct values. More precisely, each of the four
seeds of the first four MWC 16 bit generator group has been set to value 1 
(low words of x[i], i=1,..,4), each
of the four seeds of the second generator group has been set to value 2 
(low words of y[i], i=1,..,4),
and all the initial carry values have been set to 0 
(high words of x[i] and y[i], i=1,..,4).

A single update of each of the MWC 16 bit generators can be broken down in the
following elementary steps on a 32 bit block x:

\bigskip
i) extract the low word (as x$\&$0xFFFF)

ii) multiply the result with the constant multiplier

iii) extract carry from x (as x$>>$16)

iv) add results of ii) and iii)
\bigskip

\noindent 
and the MWC1616 generator
may be implemented with SSE2 instructions as follows:
\begin{verbatim}
_asm
{
movdqa  xmm2, xmm0  //make a copy of x
psrld   xmm2, 10h   //x>>16 in xmm2
//now find a*(x&0xFFFF)
//in current exemple, a=18000=0xmm1650:
andps   xmm0, xmm5  //x&0xFFFF
movdqa  xmm3,xmm0   //make a copy
pmullw  xmm0, xmm6  //multiply, save low word
pmulhuw xmm3, xmm6  //multiply, save high
pslld   xmm3, 10h   //high result << 16
orps    xmm3,xmm0   //low OR high
//a*(x&0xFFFF) now (finally) in xmm3
paddd   xmm2,xmm3   //a*(x&0xFFFF)+(x>>16)
movdqa  xmm0,xmm2   //copy new value to x
pslld   xmm2, 10h   //save x<<16 (for return)
// now second generator...
movdqa  xmm4, xmm1  //make a copy of y
psrld   xmm4, 10h   //y>>16
//find b*(x&0xFFFF), where b=30903=0x78B7:
andps   xmm1, xmm5  //y&0xFFFF
movdqa  xmm3,xmm1   //make a copy
pmullw  xmm1, xmm7  //multiply, save low word
pmulhuw xmm3, xmm7  //multiply, save high
pslld   xmm3, 10h   //high result << 16
orps    xmm3,xmm1   //low OR high
//b*(x&0xFFFF) in xmm3
paddd   xmm3,xmm4   //b*(y&0xFFFF)+(y>>16)
movdqa  xmm1,xmm3   //copy new value to y
andps   xmm3, xmm5  //and with 0FFFFh
paddd   xmm2,xmm3   //(x<<16)+(y&0xFFFF)
//random number in xmm2
movdqa  r,xmm2      //save result in array r
}
\end{verbatim}
The principal problem that was encountered when implementing
these steps in parallel, on the four adjacent 32 bit blocks stored
in the XMM registers, was to perform multiplication in step
ii). Namely, the SSE2 instruction set does not contain an adequate
instruction to multiply 16 bit values in the low words of the four
adjacent blocks (the high words are by definition zero, as both 
the current value and 
the multiplier are 16 bit integers), and store the results 
in the same 32 bit blocks. 
Instead, two distinct multiplication instructions were used for 
multiplying 16 bit blocks, and then storing first the low, 
and then the high 16 bit result, as exemplified by the 
following code snippet:
\begin{verbatim}
andps   xmm0, xmm5  //x&0xFFFF
movdqa  xmm3,xmm0   //make a copy
pmullw  xmm0, xmm6  //multiply, save low word
pmulhuw xmm3, xmm6  //multiply, save high
pslld   xmm3, 10h   //high result << 16
orps    xmm3,xmm0   //low OR high
\end{verbatim}
Here, the value (x$\&$0xFFFF) in the xmm0 register is first copied to
xmm3, the two copies are then individually multiplied by the constant
previously loaded into xmm6, 
where first low word and then high word results are
stored. Next, the high result is shifted 16 places to the left, and
finally OR-ed with the low word result, to yield the final product values.

The three different implementations of the MWC1616 generator, 
all yielding {\it exactly the same} sequences (given the same seeds) 
of uniform deviates, 
have been tested
using a Celeron 1.6GHz processor, on a Toshiba Satellite laptop computer.
The tests were performed for all three implementations, both with and
without compiler optimization (flags /Od and /O2 of the cl compiler).
Timing results (per uniform deviate) performed by averaging over 1000 blocks
of 1000000 deviates, are summarized in Tab.~\ref{tab1}.

\begin{table}[h]
%\begin{table}
\caption{\label{tab1}
Execution time in nanoseconds per MCW1616 uniform deviate, 
averaged over 20000 blocks of 50000 deviates each.
Results are presented for both debugging version,
and compiler optimized code, where $\sigma_D$ and $\sigma_O$
represent the corresponding standard deviation over the
20000 blocks.
}
%\begin{ruledtabular}
\begin{tabular}{lcrcrcrcr}
\hline
\hline
Version&&Debug&&$\sigma_D$&&Optimized&&$\sigma_O$\\
\hline
C function&&58.47&&2.94&&48.56&&3.14\\
C macro&&31.15&&2.66&&29.19&&2.40\\
SSE2&&5.72&&1.01&&5.34&&0.92\\
\hline
\hline
\end{tabular}
%\end{ruledtabular}
\end{table}

It is seen from Tab.~\ref{tab1} that compiler optimization has a
significant impact only in the case of a straightforward function
call (in the case of SSE2 compiler optimization affects only the 
outer loops, while inline assembly sequences are preserved). 
The standard deviation over the 1000 runs (of 1000000
deviates each) is presented to indicate the extent to which the
background activity of the operating system affects the execution
time. While relative fluctuations are largest in the SSE2 case,
it should be noted that a single cycle (or clock tick) of a 1.6GHz
machine has a duration of 0.625 nanoseconds: roughly ten cycles are
used per deviate, with a standard deviation of two cycles.
The fact that twenty four SSE2 assembly instructions constituting the
128 bit implementation of the MWC1616 generator are on average
executed in only ten cycles per deviate, is the result of the 
fact that four uniform deviates are calculated in parallel.
More precisely, twenty four instructions in the above code are
executed in 22.88 nanoseconds, yielding four uniform deviates.
Therefore, an effective average of one (fourfold SIMD) instruction per cycle 
is achieved even as a total of four multiplication instructions
(generally requiring more time then other instructions) and
three addition instructions were used,
which may be attributed to the pipelining
architecture of the processor \cite{intel}, representing another
aspect of the parallelism paradigm implemented by the Pentium IV processors:
instructions are processed several at a time, similar to a 
factory assembly line.

In conclusion, it follows from the above that scientific implementations
of 128 bit SSE2 extensions require some ``ugly" programming practice:
not only one has to resort to assembly language programming, but the
available instruction set is also somewhat unusual, and far from being complete.
On the other hand, the impressive 500\% gain over the optimized C code
(or for that matter straight SISD assembly) may justify such coding
practice for time critical applications, in fact, in certain situations
it may make all the difference between an unfeasible and a feasible
problem (an application that would run five months may be considered
unfeasible, and the one that would run a single month, feasible).

Finally, it should be stressed that the application to random number
generation presented in this work by no means represents the only
possible application for scientific computing, many other time critical
algorithms (e.g. linear algebra matrix manipulation) may benefit
from this approach (SSE2/3 extensions are not limited to integer arithmetic,
both single and double precision floating point algebra is supported). 
A C library with the current implementation of MWC1616
uniform deviates (which does not require
assembly programming nor installation of the processor pack, unless
one wants to recompile the library from source),
can be obtained from the author upon request.

\end{document}